\documentclass{aa}  
\usepackage{natbib}
\bibpunct{(}{)}{;}{a}{}{,} 

\usepackage{graphicx}
\usepackage{txfonts}
\usepackage{amsmath}
\usepackage{xcolor}
\usepackage{hyperref}
\hypersetup{
    colorlinks,
    linkcolor={red!50!black},
    citecolor={blue!50!black},
    urlcolor={blue!80!black}
}
\usepackage{cleveref}
\usepackage{bm}
\usepackage[normalem]{ulem}

\newcommand{\kms}{\,\rm km/s}
\newcommand{\Mpc}{\,{\rm Mpc}}
\newcommand{\Gpc}{\,{\rm Gpc}}
\newcommand{\Mpch}{\,\Mpc/h}
\newcommand{\Gpch}{\,\Gpc/h}
\newcommand{\Msh}{\,M_{\odot}/h}
\newcommand{\Hnod}{$H_0\ $}
\newcommand{\Hdim}{\kms/\!\Mpc}

\newcommand{\LCDM}{$\Lambda$CDM}

\newcommand{\ie}{i.e.,}
\newcommand{\eg}{e.g.,}
\newcommand{\param}{\pi}
\newcommand{\data}{\mathcal{D}}
\newcommand{\other}{other}
\newcommand{\muobs}{\mu^{\rm obs}}
\newcommand{\zcos}{z_{\rm cos}}
\newcommand{\st}{^{(s)}}

\newcommand{\lN}{\mathcal{N}}

\newcommand{\srH}{\sigma_{\Vr}}
\newcommand{\Vr}{\langle V_r\rangle}

\begin{document}

   \title{Prior-free cosmological parameter estimation of Cosmicflows-4}

   \author{C. Duangchan \inst{1,2}\fnmsep\thanks{Corresponding author}, A. Valade\inst{3} , N. I. Libeskind \inst{1}, \and M. Steinmetz\inst{1}
          }

   \institute{Leibniz-Institut für Astrophysik Potsdam (AIP),
              An der Sternwarte 16, 14482 Potsdam, Germany \\ \email{cduangchan@aip.de}
         \and
             Universität Potsdam, Institut für Physik und Astronomie, Karl-Liebknecht-Str. 24-25, 14476 Potsdam, Germany
        \and
            Aix Marseille Universit\'e, CNRS/IN2P3, CPPM, Marseille, France
             }

   \date{Received; accepted}

  \abstract
   {As tracers of the underlying mass distributions, the peculiar velocities of galaxies are valuable probes of the Universe, allowing us to measure the Hubble constant or to map the large-scale structure and its dynamics. The catalogs of peculiar velocities, however, are noisy, scarce, and prone to various interpretation biases.}
   {We measured the radial and bulk flow directly from the largest available sample of peculiar velocities and did not impose a cosmological prior on the velocity field. Furthermore, a minimum assumption on the shape of the radial flow at large distances enabled us to estimate the local Hubble constant.}
   {To address these issues, we analyzed the Cosmicflows-4 catalog (CF4), the most extensive catalog of galaxy peculiar velocities, reaching a redshift $z=0.1$. Specifically, we constructed a forward-modeling approach assuming only a flat Universe, which reconstructs the radial and bulk flows of the velocity field directly from measurements of peculiar velocities. Our method was tested on a series of 64 simulated catalogs that mimicked the complex selection function of CF4 in space and in magnitude. Based on our mock data, we propose a simulation-based correction method that we applied to the CF4 data.}
   {Our method recovers the radial flow and the direction and magnitude of the bulk flow throughout the covered volume without bias. The incompleteness of the data leads to a systematic amplification of the underlying flow, however, which in turn leads to a systematic overestimation of the amplitude of the bulk flow. By estimating the (cosmic) variance of the density field at large distances from the Lambda Cold Dark Matter (\LCDM{}) model, we were able to extract a value of $75.9\pm1 \,\,{\rm (stat)} \,\,\Hdim$ from the radial inflow. With regard to the bulk flow, a 3$\sigma$ tension is found with \LCDM{} on the supergalactic X direction and on the magnitude of the bulk flow around $140\Mpch$ and $240\Mpch$. In summary, our work confirms the existing tension on the Hubble constant measured locally and a significant tension in the local bulk flow with \LCDM{} predictions. More work is needed to handle all the possible sources of systematic errors inherent to the treating of composite catalogs such as CF4.}
   {}
 
   
   \maketitle
   \nolinenumbers

\section{Introduction}
\begin{figure*}[!ht]
    \centering
    \includegraphics[width=\hsize]{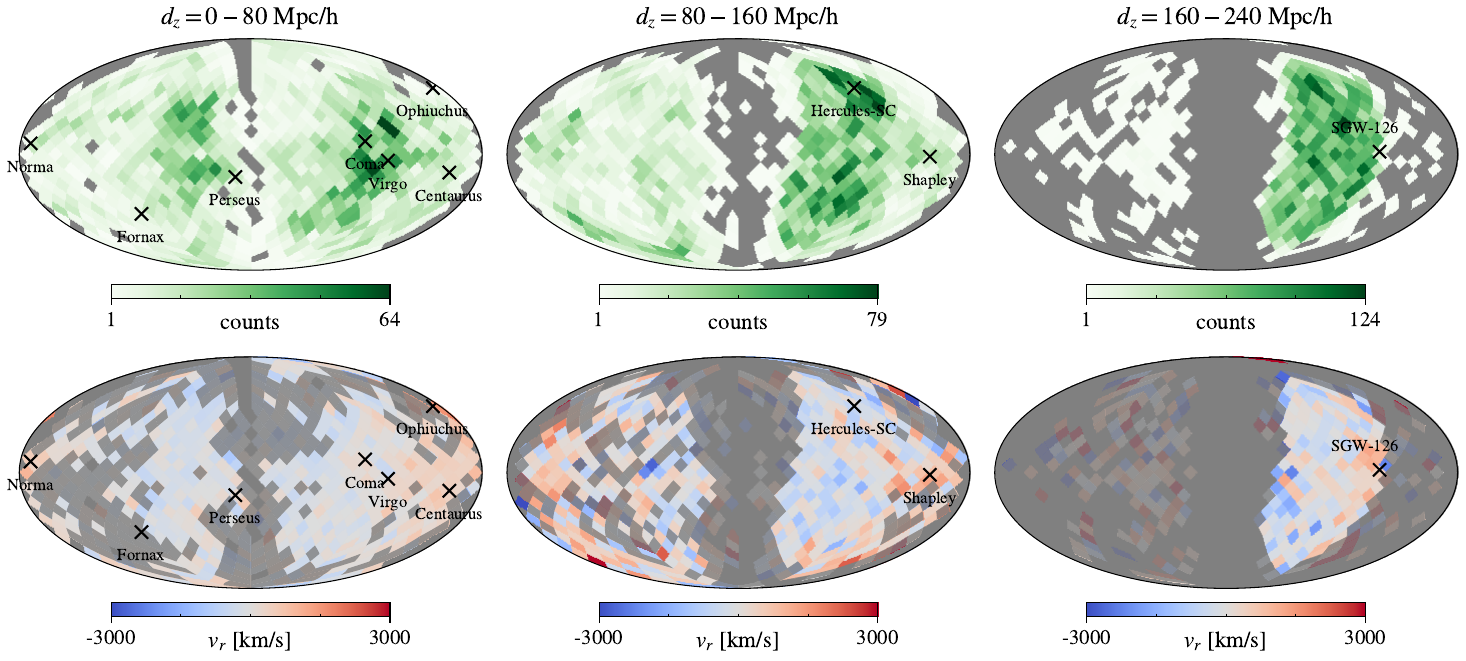} 
    \caption{Visualization of the CF4 data, split by redshift distance $d_z$. Some of the local clusters within the given range are marked with a black cross with their names. \textbf{Top:} Number of grouped galaxies within the HEALPix pixel. \textbf{Bottom:} Median of radial peculiar velocity for each direction. The pixels containing fewer than 10 data points are shaded in gray.}
    \label{fig:vrmap_sph_shl}
\end{figure*}

The motion of galaxies inspired cosmologists for at least a century when Hubble noted that the redshifting of absorption lines was proportional to distance, thereby discovering the expansion of the Universe \citep{Hubble1929}. The Hubble law, predicted by Friedman some years prior, is simple: $cz = H_{0} d$, where $H_0$ is (the Hubble) constant, $z$ is the redshift, $c$ is the speed of light, and $d$ is the distance. The redshift Hubble used, however, is now known to be a combination of the velocity due to cosmic expansion and the line-of-sight component of a galaxy's peculiar velocity, namely the velocity caused by the net gravitational pull exerted by the full mass distribution of the Universe \citep{Kaiser1987,Dekel1993}. At first order, the equation should read $cz = H_{0} d + v_{r}$, with $(v_r)$ the radial component of the peculiar velocity. Thus, the peculiar radial velocity of a galaxy can be measured based on a joint measurement of redshift and distance,
\begin{equation}
    v_r = cz - H_0 d.
    \label{eq:hubble}
\end{equation}

Catalogs combining redshifts and distances have demonstrated their relevance for cosmology. They enabled the discovery of the accelerated expansion of the Universe \citep{Riess1998, Perlmutter1999}, yielded a local value of $H_0$ \citep[see][for a review on the $H_0$ tension]{DiValentino2025}, permitted us to constrain cosmological parameters such as the growth factor $f\sigma_8$ \citep[\eg{}][]{Carreres2023, Rosselli2025} or the matter density parameter $\Omega_M$ \citep{Brout2022}, tested nonstandard models of (evolving) dark energy with cold dark matter (CDM) such as $w_0w_a$CDM \citep{Brout2022}, explored anisotropies in the expansion of the Universe \citep{Giani2024, Boubel2025, Kalbouneh2025a, Kalbouneh2025b, Stiskalek2025}, or investigated the misalignment between local motion and the Cosmic Microwave Background (CMB) dipole \citep[\eg{}][]{Sorrenti2023}. Locally, peculiar velocities have been used to map the large-scale structure \citep[\eg{}][]{Courtois2013,Valade2024} and the discovery of local dynamical structures such as the Great Attractor \citep{Dressler1987}, the Laniakea basin of attraction \citep{Tully2014}, or the dipole repeller \citep{Hoffman2017}. Last but not least, measurements of peculiar velocities can capture the so-called moments of the velocity field \citep[][namely the dipole, quadrupole, etc]{Regos1989}, which can be easily compared to theoretical predictions, providing a valuable test for cosmological models. This is the path we explore in this work.

While systematic redshift surveys with sufficient precision became available in the mid 1980s \citep{DeLapparent1986}, it remains difficult to measure the physical distances to galaxies. All distance measurements rely on essentially the same principle: the comparison between the observed magnitude or size with an assumed absolute magnitude or size. In the local Universe, where individual stars can be resolved, this is typically more accurate since (e.g.) the magnitude of stars at the tip of the red giant branch \citep{Lee1993} is a well-established standard candle. Beyond this region, absolute magnitudes need to be derived from supernovae and scaling relations such as the Tully-Fisher \citep{Tully1977} or the fundamental plane relations \citep{Dressler1987}. Supernovae, although observable to great distances, are rare, unfortunately.

For observational reasons, the vast majority of distance measurements are provided in distance moduli $\mu \propto \log(d)$. Issues arise when errors on $\mu$ (assumed to be normally distributed) are directly placed on the distance, since these then transform into log-normal distance errors, leading to biases \citep{Straus1995}. This breathing mode motivated the development of various distance estimators \citep[\eg{}][]{Watkins2015,Sorce2016,Hoffman2021, Watkins2023, Sorce2024}, which need to be applied to these data prior to any analysis,  such as the moments of the velocity field we explore here. While carefully derived, these methods remain very spontaneous and result in the designing of heterogeneous analysis pipelines.

Methods designed to reconstruct the bulk flow from corrected velocity data (i.e., data from which this log-normal bias was removed) can be separated into three families. First, the maximum likelihood method proposed by \citet{Kaiser1988} and applied to various datasets by, for example, \citet{Ma2011}, \citet{Ma2014}, \citet{Qin2018}, \citet{Qin2021}, \citet{Howlett2022}, and \citet{Whitford2023}. Second, the minimum-variance method developed by \citet{Watkins2009}, later applied by, for example, \citet{Feldman2010}, \citet{Watkins2015}, \citet{Watkins2023}, \citet{Scrimgeour2016}, \citet{Peery2018}, and \citet{Whitford2023}. Third, the Wiener filter framework, augmented by constrained realizations to sample uncertainties \citep{Hoffman1992,Zaroubi1995}, which has been applied in successive releases of the Cosmicflows catalogs (CF2, CF3, CF4) \citep{Tully2012,Tully2016,Tully2023} and associated analyses \citep{Hoffman2015,Hoffman2021,Hoffman2024}. The ability to reconstruct the full 3D velocity field of this last approach, assuming a \LCDM{} prior and the CMB-derived matter power spectrum, is among its strengths. Since it enforces a theoretical \LCDM{} velocity-velocity correlation, this approach cannot truly probe the \LCDM{} model, especially in regions beyond a few dozen megaparsec, where the noise dominates the signal and data are scarce.

An alternative approach, known as forward modeling, consists of deriving the full posterior probability of the model parameters of interest (\eg{} radial flow, bulk flow, and $H_0$) given the data using a Bayesian framework. The posterior probability is then explored with a Monte Carlo algorithm. Although computationally expensive, forward modeling surpasses more traditional methods as it can describe the complexity of the distance observations with a fidelity higher than that of conventional distance estimators. 

The application of forward modeling to reconstructing the velocity field was first proposed by \citep{Lavaux2016} and was notably applied by \citet{Graziani2019} to the Cosmicflows-3 catalog. The method has been further improved, as described by \citet{Boruah2021} and \citet{Valade2022}, and was applied to the Cosmicflows-4 data notably by \citet{Courtois2023} and \citet{Valade2024}. While the forward-modeling approach has shown a greater ability to reconstruct small scales than the traditional Wiener filter approach \citep{Valade2023}, its ability to recover the tension with \LCDM{} on the large scales is even more limited than that of the Wiener filter. In addition to assuming an underlying \LCDM{} prior, the reconstructions are performed in periodic boxes (\eg{} 1$\Gpch$ in \citet{Valade2024}), which is known to further dampen the recovered large-scale motion due to the missing large-scale tidal modes. On the one hand, strong priors tend to hide the tension in the data with respect to the model. On the other hand, in the absence of priors, parameters of interest may be less well constrained and the scientific statement is more modest.

A forward-modeling reconstruction of the first spherical modes of the velocity field has been presented by \citet{Boubel2024}, who also provided an estimate of the Hubble constant $H_0$. With a similar model, the anisotropy of the Hubble constant has been discussed by \citet{Boubel2025}. The approach developed in these works, however, focused on distances obtained from the Tully-Fisher relation, thus limiting the application of their method to the Cosmicflows-4 Tully-Fisher sample \citep{Kourkchi2022}. Although it is affected by complex magnitude-selection effects, this dataset is relatively isotropic. 

We present a forward modeling of the first two spherical modes of the velocity field (radial flow and bulk flow) from peculiar velocity data that makes minimum \LCDM{} assumptions. As it relies on the Bayesian framework, the method can directly treat the observational data without relying on an intermediary distance estimator such as \cite{Watkins2015}. Limiting the reconstruction to the first modes of the velocity enables us to depart from strong priors on the power spectrum of the velocity field, such as in \cite{Hoffman2024}, minimizing the effect of the assumed cosmology on the resulting radial and bulk flows. Finally, compared to \cite{Boubel2024}, our more generalist, but less realistic, treatment of the data allows us to apply our method to the entire Cosmicflows-4 catalog of peculiar velocities. This leads to results on the velocity flow at greater depth while raising issues of anisotropy coupled to the selection in magnitude \citep{Andersen2016a}.
 
\section{Observations}

As mentioned in the preceding section, we analyzed the radial peculiar velocity field. For this purpose, we employed the Cosmicflows-4 (CF4) catalog \citep{Tully2023}, which constitutes the most comprehensive collection of galaxy peculiar velocities to date, containing about 56\,000 galaxies gathered in about 38\,000 groups of galaxies\footnote{A group can also include a single galaxy.} so as to mitigate the effects for nonlinear motion within bound structures. This work is based on the grouped catalog, and all observed redshifts are considered in the CMB frame. 

Rather than a single survey, the CF4 catalog is an intercalibrated compilation of several subcatalogs found in the literature: a catalog of catalogs. The fundamental-plane scaling relation \citep{Djorgovski1987} provides about three-quarters of the measured distances, with an uncertainty on the distance modulus of $\sigma_\mu \sim 0.4$. Many of these data are gathered from the Sloan Digital Sky Survey Peculiar Velocity \citep[SDSS-PV]{Howlett2022} catalog, which extends to $z = 0.1$ ($d \sim 300 \Mpch$), but is restricted to the relatively narrow footprint; another significant component comes from the 6 degree field galaxy redshift survey \citep[6dFGRSv]{Magoulas2016}, which covers the entire southern celestial hemisphere up to $z = 0.055$ ($d\sim 160 \Mpch$). 

The Tully-Fisher scaling relation \citep{Tully1977} is used to derive the majority of the remaining entries, with an uncertainty similar to that of the fundamental plane. These distances are mostly provided by the CF4-TF catalog \citep{Kourkchi2022}, which is an analysis of SDSS  HI fluxes, line widths, and redshifts. CF4-TF covers roughly the entire sky, with a slight preference for the northern celestial hemisphere. The redshift distribution of the Tully-Fisher sample peaks at $z=0.02$ ($d \sim 60 \Mpch$) and slowly decreases until $z=0.055$. 

Cosmicflows-4 contains about 1\,000 type 1a supernovae extracted from the Pantheon+ and SH0ES catalogs \citep[][respecitvely]{Riess2022,Scolnic2022}, found mostly below $z=0.03$ ($d\sim120\Mpch$) and reaching $z\sim0.1$. Finally, various higher-precision distance estimators ($\sigma_\mu \leq 0.1$) are also present in the CF4 compilation, but are limited to $z< 0.02$. 

The complex footprint of CF4 can be appreciated in \cref{fig:vrmap_sph_shl}, where we show the angular distribution of the catalog, colored by the number of groups in each pixel on the sky in three different radial shells. We considered three different regimes: dense and isotropic below $80\Mpch$, relatively dense and relatively isotropic between $80-160\Mpch$, and last, strongly anisotropic between $160-240\Mpch$. The last shell was cropped to $240\Mpch$ so as to keep shells of constant thickness. The Milky Way shadows an opening angle of about $10^{\circ}$ around the Galactic plane; a region of the sky dubbed the zone of avoidance, where it is relatively impossible to obtain distance modulus measurements due to foreground obscuration.

\section{Method}
\begin{figure*}[!ht]
    \centering
    \includegraphics[width=\textwidth]{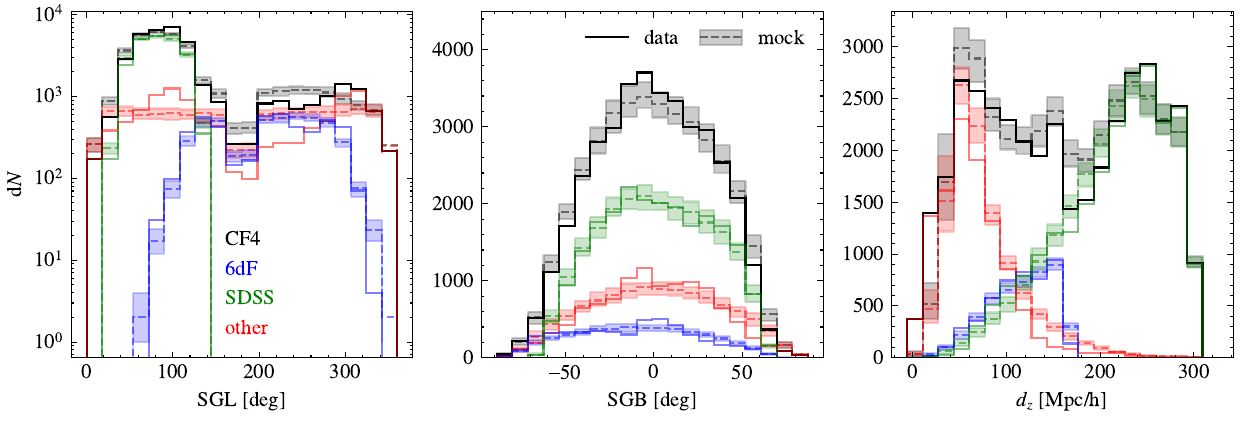}
    \caption{3D spatial distribution of the CF4 and mock catalog of each subtype shown as solid and dashed lines, respectively. The shade in a lighter color is the scatter among the mocks.}
    \label{fig:mock_spat_dist_type_1}
\end{figure*}

In this section, we detail the model and the method we used to reconstruct the moments of the velocity field from measurements of peculiar velocities. The approach can be summarized as follows. The first two moments of the velocity field (\ie{} a radial in or out flow plus a net flow in some direction) were chosen at random. Assuming that the errors in the redshift measurements are negligible, we reconstructed the distance to each galaxy from (a more accurate version of) \cref{eq:hubble} (see \cref{eq:zcos}). The inferred distances, transformed into distance moduli, were then compared with the Cosmicflows-4 observations. Finally, the moments of the velocity field were updated with a Monte Carlo step. The result is a large collection of probable values of radial and bulk flows compatible with the Cosmicflows-4 data. This approach, dubbed forward modeling, is detailed in the following section. 

\subsection{Forward modeling}

Forward modeling heavily relies on Bayes's theorem, which encodes the posterior probability of the model parameters $\param$ given that the data $\data$ were obtained as
\begin{equation}
    P(\param \mid \data) \propto P(\data \mid \param) P(\param),
    \label{eq:bayes}
\end{equation}
where the likelihood $P(\data \mid \param)$ and the prior $P(\param)$ encode the observational and the underlying physical models, respectively. As discussed in \cref{sec:dis}, we assumed no prior distribution on our parameters, that is, $P(\param) = 1$. 

We reconstructed the zeroth and first modes of the velocity field \citep{Regos1989}. These are often called the monopole and the dipole, respectively. However, since the term ``dipole'' has come to be understood in common parlance as the motion of an observer relative to background, we refer with it to the mean motion of galaxies flowing either radially (radial flow) or in some direction (bulk flow). 

Throughout this paper, the radial flow is referred to as $V_r$. Although the radial flow has the dimensions of speed (km/s) and one of two directions (inward or outward), it is not a vector. The bulk flow is a 3D vector in the traditional sense and is referred to here as $\bm V = (V_X, V_Y, V_Z)$. Our parameter space thus consists of four parameters, namely $\param = (V_r, V_X, V_Y, V_Z)$. As detailed below, these quantities fit by the forward model represent the velocity field in concentric shells, while the associated quantities in spheres were constructed as detailed in \cref{sec:shell_to_sphere}. 

We considered $\data = \{\muobs_i\}_{i < n}$, where $n$ is the number of observations (namely the number of entries in the CosmicFlows-4 catalog). The redshift does not feature in $\data$ since the error on it was neglected and therefore did not need to be inferred. Moreover, distance moduli are not per se observables, but were derived from other observables through relations such as the Tully-Fisher relation or the fundamental plane. 

\subsection{Likelihood}
\label{sec:lkl}

As mentioned above, the goal was to make a suggestion for each galaxy's (or galaxy group's) distance modulus, for the purpose of comparison with the measured value. Thus, the distance modulus for each entry was constructed as follows. First, the peculiar velocity of the constraint $i$ was computed as 
\begin{equation}
    v^{i}_r = V_r + \bm V\cdot\hat{\bm{r}}^{i}
    \label{eq:vel},
\end{equation}
where $\hat{\bm{r}}$ is a unit vector pointing toward the constraint. Assuming that the full redshift $z$ is known with negligible uncertainty, we computed the proposed cosmological redshift (namely the redshift due to the Hubble expansion) from the proposed peculiar velocity,
\begin{equation}
    \zcos = \frac{z - v_r / c}{1 + v_r / c},
    \label{eq:zcos}
\end{equation}
which is an accurate version of \cref{eq:hubble}. Then, the proposed physical distance $d$ was obtained by numerically integrating 
\begin{equation}
    \label{eq:dcosint}
    d=\frac{c}{H_0} \int_{0}^{\zcos}\frac{\text{d}z}{\sqrt{(1+z)^{3}\Omega_{\rm M}+(1-\Omega_{\rm M})}}.
\end{equation}
Last, the distance modulus was obtained,  
\begin{equation}
    \mu = 5 \log_{10}\left[(1 + \zcos) d\right] + 25. 
    \label{eq:d_to_mu}
\end{equation}

The Hubble parameter $H_0$ and the density parameter $\Omega_M$ were not explored by our forward model. Instead, their respective effects on the results are discussed in \cref{sec:results} and \cref{app:omega_m}. We note that \cref{eq:dcosint} makes assumptions on the background cosmology, which are further discussed in \cref{sec:dis}.

After these calculations were carried out, the likelihood function was constructed, namely, the ensemble of proposed values of the distance moduli were compared with the observed values and their errors. The distance modulus errors were assumed to be normally distributed around the mean value and independent of each other, which is the standard approach \citep[\eg{}][]{Hoffman2021, Boruah2021, Valade2022, Tully2023}. The likelihood over all the reconstructed distance moduli is expressed as the exponential product of the difference between the observed distance modulus and the proposed one, divided by the error on the distance modulus, 
\begin{align}
    P\left(\{\muobs_i\} \mid \param\right) & = \prod_i P\left(\muobs_i \mid \param\right) \\
                                            & \propto \prod_i \exp \left( - \frac{\left(\muobs_i - \mu_i(\param)\right)^2}{2 \sigma_{\mu, i}^2}\right).
    \label{eq:lkl}
\end{align}

We did not attempt to model the volume or selection effects (\ie{} $P(d) = {\rm constant}$). This is loosely consistent with the distance distribution of the entire CF4, which is roughly flat, as shown in \cref{fig:mock_spat_dist_type_1}.

The posterior distribution was explored by means of a Hamiltonian Monte Carlo (HMC) technique \citep{Neal2011}. An HMC is an efficient Monte Carlo technique that enables the exploration of the full permissible parameter space. Each  Monte Carlo chain comprises 1\,000 states, preceded by a 100-step burn-in phase. Given the simplicity of our problem, any reasonable choice of hyperparameters led to an optimal convergence of the algorithm. The HMC implementation made use of \textsc{TensorFlow-Probability} \citep{Dillon2017} with a \textsc{Jax} backend \citep{Bradbury2018}.

\subsection{Inference of the velocity field in concentric shells}
\label{sec:shell_to_sphere}

Since distances are measured via their observed distance moduli, the error on the peculiar velocity $(\sigma_v)$ for each entry of the catalog is proportional to the measured error on the distance modulus ($\sigma_{\mu}$) and its (true) distance. For the vast majority of data points, the redshift distance is more precise than the physical distance inferred from the distance ladder,
\begin{equation}
    \sigma_v \approx \frac{\log(10)}{5} \sigma_\mu \times cz. 
    \label{eq:sig_v}
\end{equation}

Owing to distance-dependent errors, most of the constraining power of the data is found near the observer. This implies that running the inference method on the entire catalog at once would lead to a bulk flow estimated at an effective radius
\begin{equation}
    r_{\rm eff} = \frac{c}{H_0} \frac{\sum_i z_i / \sigma_{v, i}}{\sum_i 1 / \sigma_{v, i}} \approx 100 \Mpc~~(\approx 75 \Mpch),
    \label{eq:eff_rad}
\end{equation}
which is much shallower than that of the data (whose median distance is 155\Mpch). 
This is due to (1) the decrease in the number of constraints with distance and (2) the increase in individual error on the constraints. Additionally, it is to be noted that the CF4 catalog becomes very anisotropic beyond $165\Mpch$ ($z=0.055$).

We designed an approach to overcome this limitation. Essentially, the Monte Carlo fits were first run independently in concentric shells, and then, the values in concentric spheres were reconstructed by averaging the contributions of the shells by their respective volume. More precisely, the catalog was split (according to each galaxy redshift distance $r_{i} \approx cz_{i}/H_{100} \Mpch$) into concentric shells of thickness $\Delta r = 20 \Mpch$, in each of which the Monte Carlo algorithm was run separately. The value of the sphere of radius $(R)$ was then computed as
\begin{equation}
    \label{eq:volume}
    \begin{aligned}
        \quad r_j &= j \times \Delta r,\\
        \mathcal{V}_j &= \frac{4}{3} \pi  \left(r_j^3 - r_{j - 1}^3\right),\\
        Q\st(R) &= \frac{\sum_{j}^{r_j < R} Q_j\st \mathcal{V}_j}{\sum_{j}^{r_j < R} \mathcal{V}_j}, 
    \end{aligned}
\end{equation}
where $Q_j\st$ is the $s$th Monte Carlo step of the quantity $Q$ in the $j$th shell (\ie $Q_j$). \Cref{eq:volume} was applied independently to the radial flow ($V_r$) and the components of the bulk flow $(V_X,V_Y,V_Z)$.

$Q\st(R)$ only constitute a pseudo-Monte Carlo chain and not an actual radius-dependent reconstruction. Namely, correlations between fitted values at different radii cannot be interpreted. 

\subsection{Mock of CF4}
\label{method:mocks}

\begin{table}
    \caption{Summary of the properties of each subcomponent of the CF4 mock catalogs.}\label{tab:mocks}
    \begin{center}
        \begin{tabular}[c]{l|c c c c } 
            Component & N & $\beta$ & $m_c$ & $\sigma_\mu$ \\
            \hline
            SDSS-PV & 22\,467 & 0.6 & 13.3 & 0.5 \\
            6dFGRSv & 4\,768 & 1.5 & 11.7 & 0.5 \\
            \other & 10\,822 & 0.45 & 10 & $\mathcal{N}(0.4, 0.1)$ \\
        \end{tabular}
    \end{center}
    \tablefoot{Here, N is the number of galaxies, $\beta$ and $m_c$ are the parameters fit to reproduce a magnitude-limited sample following the procedure given in \cref{app:magcut}, and $\sigma_{\mu}$ is the typical error on the distance modulus. $\mathcal{N}$ represents a normal distribution.}
\end{table}

Before the forward-model method described above was applied to the CF4 data, we determined its precision and accuracy, namely how well it measures the two modes on the velocity field in which we were interested (the radial infall and the bulk flow). It was thus applied to mock observations of the CF4 grouped catalog. In this way, the estimated velocity field can be compared directly to the ground truth. Specifically, due to its complex footprint (anisotropy, sparse sampling, large errors, and luminosity limitation), the construction of reliable mocks of the CF catalog is nontrivial, and several methods have been developed \citep[\eg{}][]{Doumler2013,Sorce2016,Qin2021,Valade2023,Whitford2023,Boubel2024}. We note that applying the method to mock catalogs drawn from simulations necessitates choosing a cosmology. In this way, the accuracy of the method loses its prior-free nature. However, we recall that the assumption of a cosmology only affects the interpretation of the results: the method itself remains largely cosmology-free.

The \textsc{Big MultiDark} $N$-body simulation \citep{bigmd}, which covers a volume of $L_{\rm box}=(2.5 \Gpch)^3$ with $N=3840^3$ particles, was used for this purpose. The cosmology of the simulation is similar to that of Planck XIII \citep[$h=0.6777$, $\Omega_M=0.307$;][]{PlanckCollaboration2016}. Halos with more than 20 particles were considered, namely a lower-mass resolution of $4.7\times10^{11} \Msh$. 

The cosmological box was sliced into $4\times4\times4=64$ cubic subboxes of side length $(625\Mpch)^3$, each of which was sufficiently large to enclose the full CF4 catalog. The observer was set at the center of each subbox. We acknowledge that this simplistic approach implies that the observers are statistically found in a cosmological underdense region, which affects results on the radial flow as discussed in \cref{sec:mock_rad_flow}. The box coordinates were (arbitrarily) associated with supergalactic coordinates (SGX, SGY, and SGZ). 

To grasp the complexity of the CF4 dataset, we separately constructed three of its subcatalogs: SDSS-PV, 6dFGRSv, and ``other''. This enabled us to model the anisotropic footprints of the scaling-relation surveys contributing to CF4, since the catalog anisotropy is dominated by the two wide-angle sky surveys (SDSS-PV and 6dFGRSv). 

Our method also focused on modeling a (simple) magnitude-limitation\footnote{This effect is also often referred to as a Malmquist bias. We refrain from using this term, as it has been used for a wider range of observational biases.} . The large number of instruments and distance-derivation methods on which the CF4 catalog is based makes a magnitude analysis very difficult. Not only are CF4 entries extracted from surveys with different magnitude limitations (among other method-dependent selection effects), but their magnitudes are observed in different bands. Furthermore, the simulation from which the halos are taken is dark-matter only, meaning that neither are halos populated by galaxies, nor are their luminosities simulated. We thus opted for an empirical approach detailed in \cref{app:magcut}. A more accurate procedure is left to further works.  

For each subbox, CF4-mock halos were extracted from the simulation according to the following recipe.
\begin{enumerate}
    \item An angular mask was applied that mimicked the sky coverage of the surveys: the footprint of the SDSS-PV, the southern celestial hemisphere (dec<$0^\circ$) for 6dFGRSv, and the full sky for other. We also included an additional cut of $10^{\circ}$ around the Galactic plane so as to model the zone of avoidance. Halos were included or excluded based on whether they were inside or outside of the survey footprints and the zone of avoidance.
    \item Each halo was given an observed redshift and distance modulus by applying \cref{eq:zcos,eq:dcosint,eq:d_to_mu} to the radial peculiar velocity and distance of each halo. 
    \item Galaxies were painted onto halos by computing an absolute magnitude from the halo mass. For each subcatalog, the empirical luminosity-mass relation was fit alongside an upper (apparent) magnitude cut so as to reconstruct the radial distribution of the observations. The exact procedure is detailed in \cref{app:magcut}. 
    \item Halos were then randomly selected out of the magnitude-limited sample.
    \item An observational scatter depending on the subcatalog was added on the distance moduli following \cref{tab:mocks}.
\end{enumerate}
The mock CF4 catalog was constructed by merging the mock SDSS-PV, 6dFGRSz, and the \other  data. 

In \cref{fig:mock_spat_dist_type_1} we show the angular and redshift distributions of the mock observations and the CF4 catalog. The spatial distribution of the CF4 catalog is reproduced with great fidelity, demonstrating the ability of our mock procedure to reproduce the complexity of the CF4 compilation well.

While the distributions of SDSS-PV and 6dFGRSv agree well in \cref{fig:mock_spat_dist_type_1}, some disparities can be seen in the \other{} component. Although the distribution in $d_{z}$ is well captured, the angular distributions (SGB and SGL) have minor differences. This is due to the crude modeling of the \other{} component, which in reality contains a number of different anisotropic surveys.

\begin{figure}
    \includegraphics[width=\hsize]{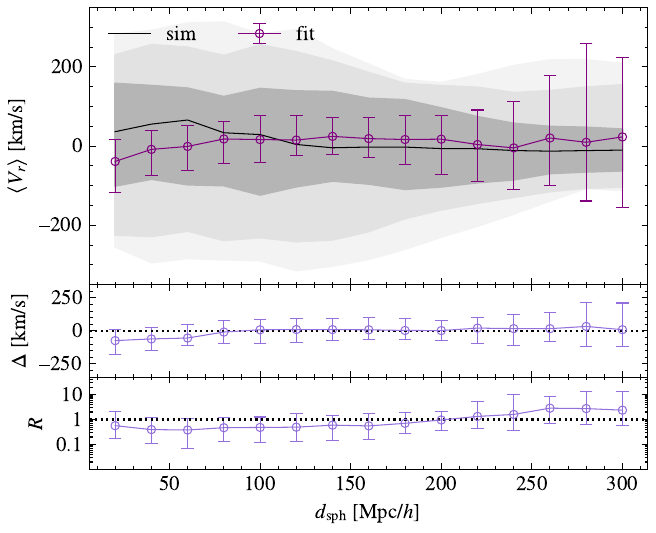}
    \caption{Comparison between the average radial peculiar velocity from the simulation and the fit results. In the top panel, the black line represents the median of the true value across 64 mocks, with the shaded gray regions indicating 68\%-95\%-99.7\% percentile intervals, equivalent to 1, 2, and 3$\sigma$. The purple error bars represent the median of the fit results, with 68\% error bars. The middle and bottom rows show the difference $\Delta$ and ratio $R$ between the estimated and simulation values, respectively. For each of these parameters, the median across the 64 mocks is shown by the central light purple line, with error bars indicating the percentile interval covering the central 68\% of the distribution.
}
    \label{fig:mock-monopole}
\end{figure}
\begin{figure*}
    \centering
    \includegraphics[width=\textwidth]{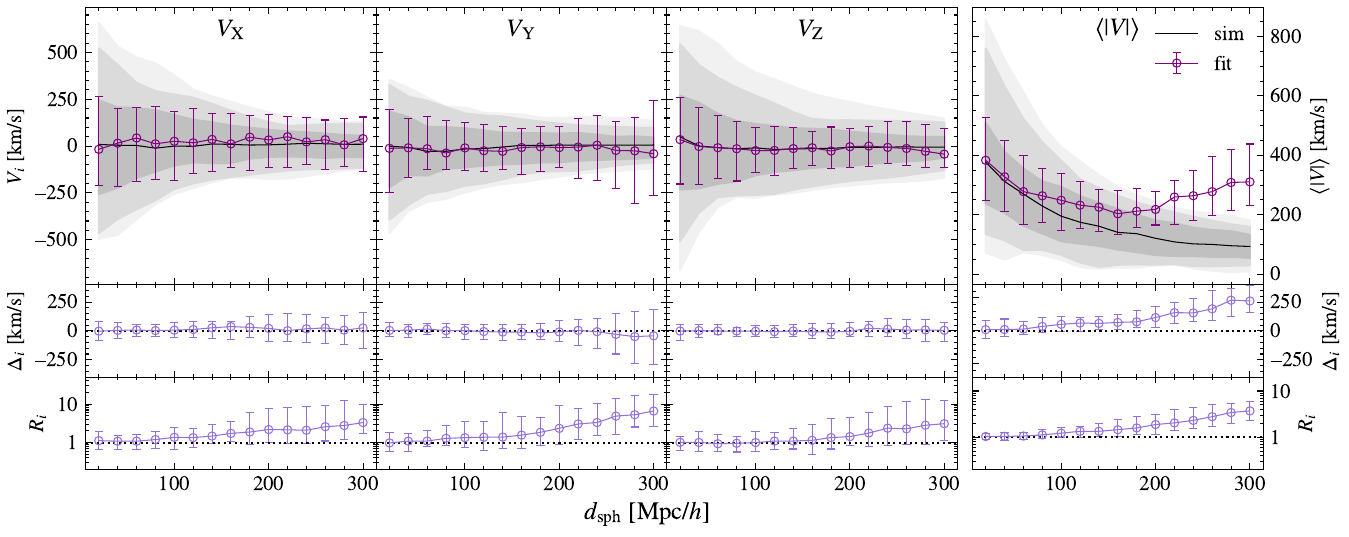}
    \caption{Same as \cref{fig:mock-monopole} but for the three Cartesian components of the bulk flow ($V_{X/Y/Z}$, first three columns) and its magnitude (right-most panel).}
    \label{fig:mock-dip}
\end{figure*}

\section{Results: Mock data}
\label{sec:results}
\begin{figure}
    \includegraphics[width=\hsize]{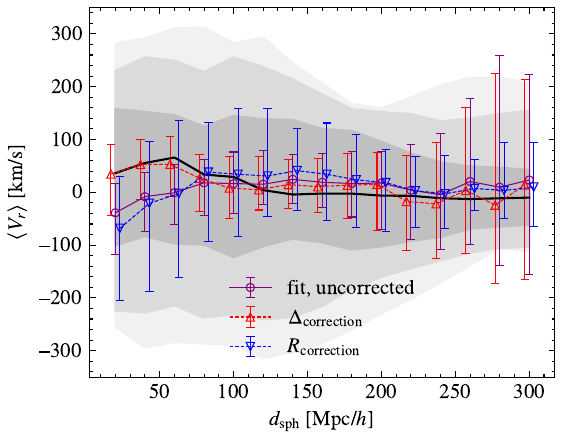}
    \caption{Application of the corrections to the radial flow fit of the mocks. The median of the radial flow in the simulation, expected to be minimal, among the mocks is shown as a black line, with gray shades indicating the 68\%-95\%-99.7\% intervals demonstrating the cosmic variance. Corrections based on differences and ratios are shown as dashed red and blue lines, respectively, with error bars indicating the 68\% interval across the 64 corrected radial flows.}    
    \label{fig:mock-mon-corr}
\end{figure}

\begin{figure*}
    \centering
    \includegraphics[width=\textwidth]{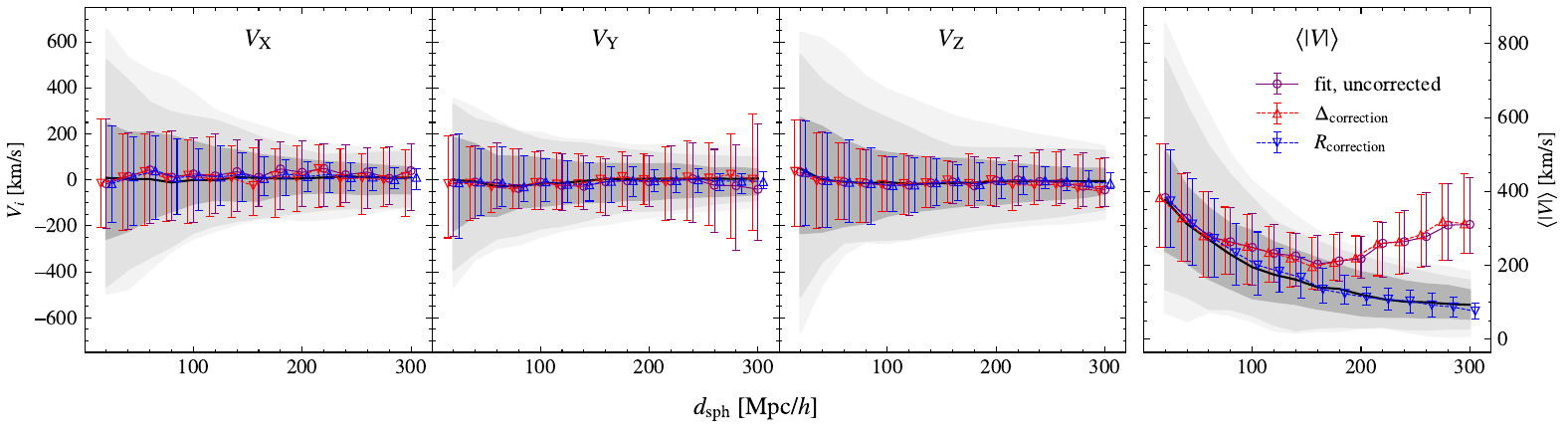}
    \caption{Same as \cref{fig:mock-mon-corr} but for the bulk flow components and magnitude. Note that corrections were applied to the components, not to the magnitude of the vector.}    
    \label{fig:mock-dip-corr}
\end{figure*}

\subsection{Radial flow}
\label{sec:mock_rad_flow}
The reconstruction of the radial flow of the 64 mocks in spherical volumes out to $300\Mpch$ is presented in the top panel of \cref{fig:mock-monopole}. The recovered radial flow closely follows the ground truth of the simulations, demonstrating that our method reliably reconstructs the radial flow within the expected statistical uncertainty. The error in the method due to the sampling increases with distance, with a remarkable change in behavior after $160 \Mpch$. As expected, this effect stems from the observational limitations modeled in our mock data, namely scarcity of the data, large observational uncertainties, and luminosity limitations. 

In the bottom panels of \cref{fig:mock-monopole}, the under- or overestimate of the radial flow as a function of distance is shown in two ways. We focus on the misestimate in km/s represented by $\Delta=V_{r}-V^{\rm sim}_{r}$, namely the estimated infall velocity at a given distance minus the corresponding value from the simulation. We also examined the ratio $R =  |V_{r}| / |V^{\rm sim}_{r}|$. The method underestimates the slight outflow found below $50\Mpch$ seen in the simulations. The rising of the median $\langle V_{r}\rangle$ within this region reflects the observer's underdense position. The underestimation of this outflow is due to the (poor) simulation resolution (in the exact region where the observations can see the faintest objects) and implies a disparity between the CF4 catalog and our mocks in the innermost region (also seen in \cref{fig:mock_spat_dist_type_1}). In the absence of constraints, the mean tends to zero while the scatter increases, but due to the absence of a \LCDM{} prior, this convergence is expected to be weaker than it is for field-level inference methods such as \citet{Hoffman2024}.

On larger scales, beyond $\sim 200\Mpch$, the method slightly underestimates the true infall velocity by about $20\kms$, which is small with respect to the expected \LCDM{} cosmic variance (more than $100\kms$ at $200\Mpch$). More prominently, observational limitations amplify the signal starting $160\Mpch$, reaching a factor of two at the $200\Mpch$. The reconstructed velocity and true velocity begin to diverge at a distance of about $160\Mpch$. This result is expected, as CF4 becomes strongly anisotropic beyond this distance, as it is then limited to the SDSS-PV sample, which covers only a small fraction of the sky.

\subsection{Bulk flow}

How well the forward model proposed here works is shown in the bottom two rows of \cref{fig:mock-dip}. Out to $\sim140\Mpch$, where the CF4 catalog is relatively isotropic, the method is able to recover each component with a variance comparable to that of \LCDM{}. No significant over- or underestimation is found in any of the components. Beyond $160\Mpch$, where only the SDSS-PV dominates, the method amplifies the underlying signal, leading to errors that exceed cosmic variance and reach a factor of 5 at the edge of the catalog. We note that the $R$ factor increases faster on the Y component than on X and Z, which we ascribe to the geometry of the survey: the SDSS-PV is concentrated around the Y-axis.

The magnitude of the bulk flow (namely the square root of the sum of the squares of each component) is presented in the right-most panel of \cref{fig:mock-dip}. We show the direct application of \cref{eq:volume} to the Monte Carlo computed magnitude $Q\st = |V|\st$ (in purple). The estimated magnitude of the bulk flow is overestimated at all distances: the greater the distance, the greater the overestimation. We distinguish three regimes: (1) up to $60\Mpch$, the reconstructed amplitude is accurate, (2) between $60\Mpch$ and $160\Mpch$, the bulk flow decreases more slowly than the \LCDM{} predictions; it is characterized by a slow, distance-increasing amplification, and (3) beyond $160\Mpch$, the amplification becomes stronger and the reconstructed bulk flow begins to increase artificially, which is at odds with the \LCDM{} prediction. 

The salient points to be gleaned from \cref{fig:mock-dip} are thus that
\begin{itemize}
    \item the components and the amplitude of the bulk flow are well estimated within $\sim 60 \Mpch$;
    \item a modest amplification is found on all components, mostly Y, between $60 \Mpch$ and $160\Mpch$, resulting in a noticeable overestimation of the amplitude of the bulk flow in this distance range;
    \item a strong amplification of each of the components and thus of the amplitude of the bulk flow beyond $160\Mpch$ appears to be symptomatic of the strong anisotropy of CF4 in this region. 
\end{itemize} 

This analysis shows that the CF4 observational limits affect the reconstructed bulk flow and lead to an artificial amplification of the estimated bulk velocity. Thus, a naive interpretation of the reconstructed bulk flow would result in an important but artificial tension with the cosmological model. 

\subsection{Simulation-based correction}
\label{sec:corr}

The tests on mock catalogs conducted in the preceding section not only give us confidence in the regime where we expect the method to perform well, but also indicate how we can interpret (and thus correct for) the regions in which the method overestimates the radial or bulk flow. In each shell of each component of the bulk flow and separately for the radial flow, we can use the value of $R$ and $\Delta$ to correct the measured value and thus recover the true velocity field moments. This was done by
\begin{enumerate}
    \item removing the mean over- or underestimation, namely subtracting the value of $\Delta$, (the $\Delta$ correction);
    \item dividing by a factor of the amplification $R$ ($R$ correction).
\end{enumerate}

The correction we applied was based on a set of mocks, and there is an obvious variance across these mocks and their corrections. In other words, in some of our mocks, the method might perform well, while in others, it might perform worse. In general, we took the median correction, but we also included this variance in the error bar of our corrected results.

To account for the variance stemming from the correction, we write 
\begin{align}
    \label{eq:D_corr_err}
    P_{\Delta}(Q_{\rm corr} \mid D) &= \int P(Q_{\rm fit} = Q_{\rm corr} - \Delta \mid D) P(\Delta) {\rm d} \Delta,\\
    \label{eq:R_corr_err}
    P_R(Q_{\rm corr} \mid D) &= \int P(Q_{\rm fit} = Q_{\rm corr} \times R \mid D) P(R) {\rm d},R 
\end{align}
where $P(Q_{\rm fit} \mid D)$ is the result of the forward model described above, and $P(\Delta)$ and $P(R)$ are the probability distributions of the correction factors. In practice, the uncorrected output of the forward model was redressed separately by each of the 64 $R$ or $\Delta$ corrections derived from the 64 mock universes, so as to sample the uncertainty in the correction arising from cosmic variance. The final results consist thus of 64 corrected profiles, on which the median and quantiles were computed. We note that this method might overestimate the uncertainty.
 
As shown in \cref{fig:mock-mon-corr,fig:mock-dip-corr}, the $R$ and $\Delta$ corrections behave differently depending on whether the bulk flow or the radial flow is examined: the $\Delta$ correction is better suited to radial measurements, while the $R$ correction works ideally for the bulk flow. We highlight that the correction on the bulk flow was performed on the components, not on the amplitude. 

\section{Results: Cosmicflows-4 data}\label{sec:result_cf4}
\begin{figure}
    \centering
    \includegraphics[width=\hsize]{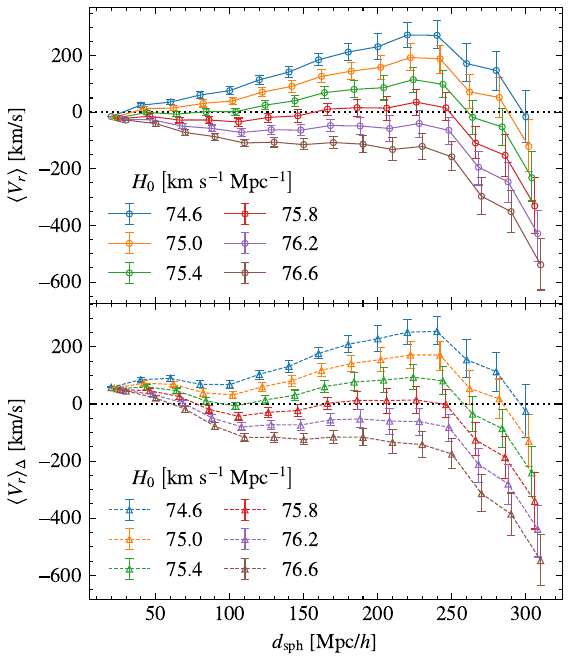}
    \caption{Radial flow fit of the CF4 data. The top panel shows the result directly from the forward modeling as solid lines, and the bottom panel shows the $\Delta$-corrected result as dashed lines. Each line, corresponding to a different value of \Hnod assumed in the model, is distinguished by color. The error bars are the 68\% interval of the Monte Carlo steps.}
    \label{fig:fit-CF4-mon}
\end{figure}

\subsection{Hubble parameter and radial flow}

Catalogs of calibrated distances, such as CF4, are the best way to obtain a local estimate of the Hubble parameter $H_0$. A simple regression of the $cz = H_0 d$ scatter plot might be run, but this is prone to several issues. First, the fact that most of the information is concentrated near the observer (where the observations are dense and uncertainties are small) leads to a very small effective radius of the fitting. Second, this approach overlooks the large-scale radial modes of the velocity. Furthermore, in the presence of an anisotropic sampling of the sky, such as CF4, the bulk flow might pollute the estimate of $H_0$ and so might higher modes of the velocity field, although observations tend to show that they have significantly less power \citep{Kalbouneh2025b}. 

We introduce a more elaborate approach. For several values of $H_0$, the radial component of the velocity flow was fit to the data. The corresponding profiles are presented in \cref{fig:fit-CF4-mon}, where the uncorrected (top) and $\Delta$ corrected (bottom) mean radial flow is presented. Two regimes appear. Up to $200\Mpch$, the velocity profiles are straight, with a slope depending on the value of $H_0$. This is the expected behavior in the presence of an offset in the value of $H_0$. Beyond $200\Mpch$, all the profiles plummet. This can be explained by the combination of (1) the limitation in the sky of the coverage in the SDSS-PV volume, and (2) the overwhelming effect of the Sloan Great Wall \citep{Gott2005,Valade2024} in this region, whose distance coincides with the elbow in the radial flow (\ie{} about $240\Mpch$). In essence, all entries in the CF4 catalog beyond $200\Mpch$ fall onto this massive object, creating a local positive radial flow between $200\Mpch$ and $240\Mpch$ and a negative radial flow beyond this distance. A correction of the radial flow changes the local inflow, as indicated in \cref{fig:mock-monopole}, and the method underestimates the mean radial flow below $100\Mpch$.

In the absence of prior information on the radial flow, all the values of $H_0$ are equally likely. In other words, without knowing on which scale no radial flow is expected, it is impossible to further constrain the expansion rate. This scale is a prediction of any cosmological model (\eg{} a consequence of the power spectrum of initial perturbations). In the \LCDM{} model, $\Vr$ is driven to zero with increasing distance according to the Copernican principle. Symptomatic of the so-called cosmic variance, the progressive convergence to homogeneity can be described by a radius-dependent probability law on $\Vr$. This allowed us to constrain $H_0$ at each distance probabilistically. Our method essentially reconstructs the probability of $H_0$ at a certain distance by marginalizing the fitted radial flows over the values predicted by cosmic variance. 

The expected \LCDM{} $\Vr$ is a Gaussian distribution,  namely  $P\left(\Vr\right) = \lN(0, \srH)$. That this is centered on zero reflects the cosmological principle of homogeneity, while $\srH$ is the scatter in the radial flow due to cosmic variance. $\srH$ can be analytically derived assuming a power spectrum or empirically sampled from a \LCDM{} simulation. In the scope of the scope of the linear theory of structure formation, $\Vr$ and $\srH$ both decrease as one over distance\footnote{In a uniform density field superimposed with a power spectrum of fluctuations, the radial velocity as a function of distance is $v_{r}(r) = -\frac{1}{3} H_0 f r \bar{\delta}(r)$, where $f$ is the linear growth factor, and $\bar{\delta}$ is the density contrast. The scatter depends directly on the scatter in the density field: $\srH^2(r) = \left( \frac{H f r}{3} \right)^2 \sigma_\delta^2(r)$}.

The probability of a given $H_0$ given the measured $\langle V_{r}\rangle$ and the \LCDM{} prior assumption of the scatter $\srH$ is 
\begin{align}
    P(H_0 \mid {\rm CF4}, \srH) & = \int P(H_0 \mid \Vr, {\rm CF4})\ P(\Vr \mid \srH)\ {\rm d} \Vr \\
    & = \int P(\Vr \mid H_0, {\rm CF4})\ \lN(0, \srH)\ {\rm d} \Vr.
    \label{eq:H0_int}
\end{align}
We considered that $P(H_0\mid \Vr, {\rm CF4})$ was obtained from the forward model of \cref{sec:lkl} and does not depend on the cosmological model\footnote{Which means here that the forward model does not depend on $\srH$, while it is true that the cosmological model enters the redshift-distance relation \cref{eq:dcosint}; see the discussion in \cref{sec:dis}.}. The equality $P(\Vr \mid H_0, {\rm CF4}) = P(H_0 \mid \Vr, {\rm CF4})$ is the application of the Bayes theorem in the absence of a prior on either $H_0$ or $\Vr$, which is consistent with the absence of a prior in the forward model. The quantity $P(\Vr \mid H_0, {\rm CF4})$ is effectively the posterior of the forward model described above, whose results are shown in \cref{fig:fit-CF4-mon} at various distances. The integral \cref{eq:H0_int} was computed numerically. 

\begin{figure}
    \centering
    \includegraphics[width=\hsize]{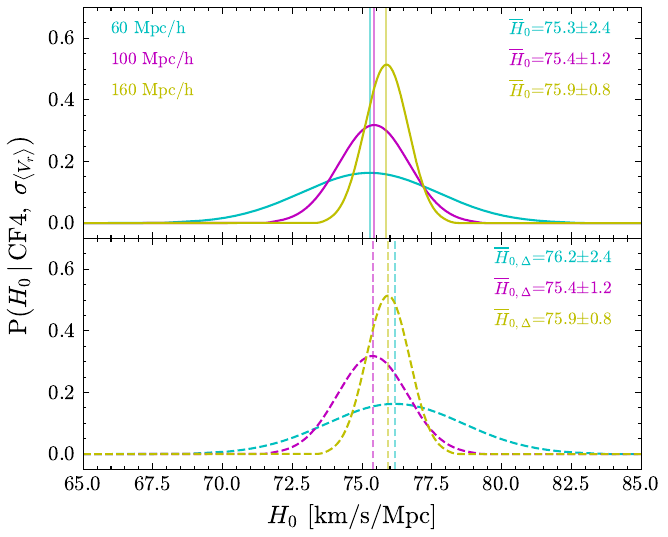}
    \caption{Probability distribution of the Hubble constant given the CF4 data and \LCDM{} radial flow variance at different sphere radii shown in colors, with $\mu\pm1\sigma$ of each curve. The vertical lines emphasize the curve mean. The marginalized likelihood function is proportional to the (un-) corrected radial flow distribution, whose resulting distributions are shown in the (top) bottom panel.}
    \label{fig:H0_pdf}
\end{figure}

\Cref{fig:H0_pdf} shows the resulting distribution of $H_0$ for the corrected and uncorrected radial flow estimator at several distances. The distributions appear to be close to normal, and the interpretation is therefore straightforward. The corrected and uncorrected values agree except at the smallest distance ($60\Mpch$). As they are computed in concentric spheres, the values of $H_0$ at different distances are not independent. To summarize our results, we chose $160\Mpch$ as our fiducial distance. This is the largest distance we trusted — it lies at the edge of the 6dF data, outside the volume dominated by the Sloan Great Wall, and appears insensitive to the correction (see \cref{fig:mock-mon-corr}).
The final result thus reads $H_0 = 75.9 \pm 1 \Hdim$. We highlight that the uncertainty might increase strongly with a proper treatment of further systematics errors in the data and in the approach\footnote{\citet{Boubel2024} reported a systematic error of $3\Hdim$ for the CF4 Tully-Fisher sample.}.

\subsection{Bulk flow}

The $V_{Y}$ and $V_{Z}$ components of the bulk flow are generally well behaved and fully consistent with the \LCDM{} predictions, with and without the $R$ correction. This is seen since the magenta and blue curves are within the 3$\sigma$ \LCDM{} expectation for essentially all distances, with a tension below $1\sigma$ before $160\Mpch$. Beyond 160$\Mpch$, the $R$-correction brings the bulk flow components in line with \LCDM{} expectations and alleviates the moderate tension found in the direct results of the forward model. 

This is markedly not the case for $V_X$, which exits the $3\sigma$ range of \LCDM{} at a distance of $\sim140\Mpch$, reaching a maximal tension with the cosmological paradigm at more $4\sigma$ level at $240\Mpch$ for the corrected and about $8\sigma$ at $240\Mpch$ for the uncorrected curves. Although the uncorrected case looks fully inconsistent with the \LCDM{} model in the region $140-300\Mpch$, the corrected $V_X$ is marginally consistent (\ie{} at the 3$\sigma$ level) with \LCDM{} between $140-220\Mpch$. 

\begin{figure*}
    \includegraphics[width=\hsize]{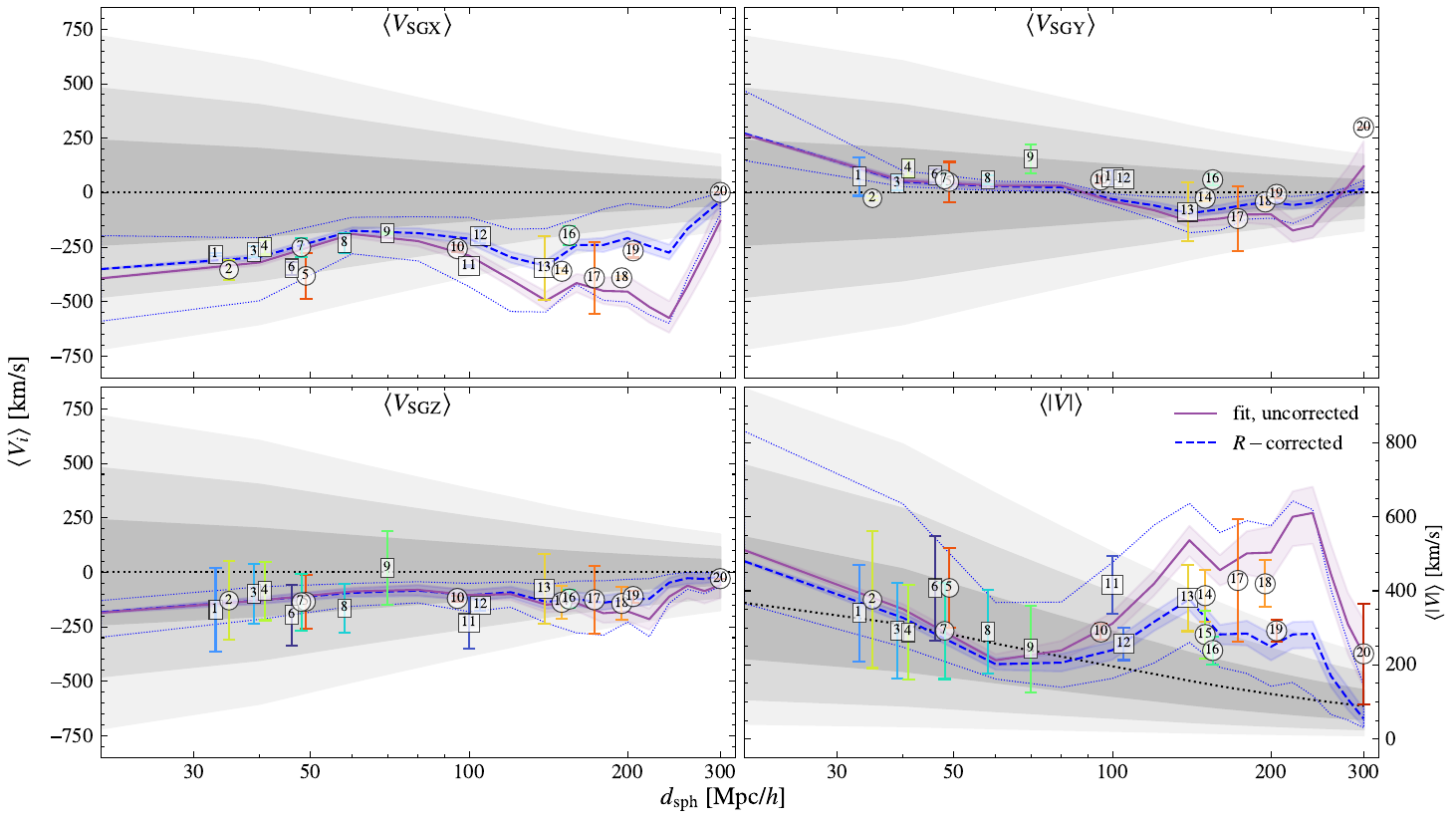}
    \caption{Bulk flow of the CF4 catalog, given $H_0=75.9$ km/s/Mpc. The purple solid line shows the direct application of our method with its $1\sigma$ uncertainty band, and the dashed blue line shows the $R$-corrected results presented in \cref{sec:corr}. The color shaded region corresponds to the uncertainty stemming from the forward model while the dotted blue lines delimit the corridor of the uncertainty coming from the $R$ correction described in \cref{eq:R_corr_err}. Other reported values of the bulk flow are denoted by numbers referring to \cref{tab:ref-bulkflow}, with values derived from the Cosmicflows series shown as circles and those from other surveys shown as squares. The radius for some literature values is approximate for illustrative purposes; the exact values are provided in \cref{tab:ref-bulkflow}.}
    \label{fig:summary}
\end{figure*}

We now analyze the amplitude of the bulk flow, which we show in \cref{fig:summary}. The direct application of the method and the corrected bulk flow decrease with increasing radius for the innermost $\sim60\Mpch$ . This behavior is typical of \LCDM. Beyond this distance, however, the CF4 data display a rise which, in the case of the corrected flow, peaks at $350\kms$ at $\sim 140\Mpch$. This value is well within the $3\sigma$ scatter of the \LCDM{} model. We assume that this behavior is due to the discrepant $V_X$ component. The  (uncorrected) bulk flow, on the other hand, exploded and reached a peak of more than $600\kms$ at $240\Mpch$.

We summarize the results of the bulk flow by noting the following: while a rise in one component ($V_X$) and in the amplitude of the bulk flow is rare in \LCDM{}, the result presented here is consistent with all other published analyses of CF4, and indeed with a few other studies that were based on other methods and data. \Cref{fig:summary} presents the values of bulk flow found in the literature at various distances, using various methods. Our approach yields results that are consistent with the literature and the \LCDM{} model for the regime out to $60\Mpch$, where the direct reconstruction of the bulk flow appears to be reliable, and beyond $60\Mpch$, where the simulation-based correction is effective. The rise in the amplitude of the bulk flow is unusual but appears to be real, or is at least a feature of the currently available data. Some methods even measured even higher bulk flows than presented here, specifically when compared to the corrected bulk flow. 

\begin{table*}
    \caption{Literature values of bulk flow from different methods and data, grouped for every 50 Mpc/h interval by horizontal lines.}
    \centering
    \small
    \begin{tabular}{clccccc}
    \hline
    No. & authors & $d_{\rm sph}$ [Mpc/$h$] & $B_{\rm SGX}$ [km/s] & $B_{\rm SGY}$ [km/s] & $B_{\rm SGZ}$ [km/s] & $|B|$ [km/s]\\ \hline
    1 & \citet{Ma2011} & 33 & -283.9 & 73.2 & -172.2 & 340 \\
    2 & \citet{Qin2021} & 35 & -353.7 & -25.7 & -129.9 & 376 \\
    3 & \citet{Hong2014} & 40 & -271.4 & 44.3 & -99.2 & 292.3 \\
    4 & \citet{Qin2018} & 40 & -250.4 & 113 & -86 & 288 \\
    5 & \citet{Whitford2023} & 49 & -382 & 48 & -135 & 408 \\
    6 & \citet{Watkins2009} & 50 & -347.7 & 78.7 & -196.3 & 407 \\
    7 & \citet{Watkins2015} & 50 & -251 & 60 & -136.3 & 292 \\ \hline
    8 & \citet{Ma2014} & 58 & -231.2 & 58.9 & -164.8 & 290 \\
    9 & \citet{Scrimgeour2016} & 70 & -186.1 & 155 & 19.2 & 243 \\
    10 & \citet{Courtois2025} & 100 & -257 & 58 & -120 & 289 \\
    11 & \citet{Feldman2010} & 100 & -337.4 & 69.6 & -233.2 & 416 \\
    12 & \citet{Nusser2011} & 100 & -198 & 61 & -151 & 257 \\ \hline
    13 & \citet{Howlett2022} & 139 & -345 & -88 & -75 & 381 \\
    14 & \citet{Watkins2023} & 150 & -360.8 & -25.5 & -137.6 & 387 \\
    15 & \citet{Peery2018} & 150 &  &  &  & 282 \\
    16 & \citet{Hoffman2015} & 150 & -196 & 59 & -123 & 239 \\ \hline
    17 & \citet{Whitford2023} & 173 & -391 & -119 & -126 & 428 \\
    18 & \citet{Watkins2023} & 200 & -391 & -43 & -143 & 419 \\
    19 & \citet{Hoffman2024} & 200 & -267 & -8 & -111 & 292 \\ \hline
    20 & \citet{Courtois2023} & 300 & 1 & 299 & -29 & 230 \\ \hline
    \end{tabular}
    \label{tab:ref-bulkflow}
    \tablefoot{Entry numbers refers to \cref{fig:summary}.}
\end{table*}

\section{Discussion}
\label{sec:dis}
We have explored a forward-modeling approach to jointly reconstruct the first two modes of the velocity field from measurements of peculiar velocities: the radial inflow, and the bulk flow. The forward model was fit independently on the data, which were split into concentric shells before they were reassembled in spherical volumes, each shell being weighed by its volume. This enabled us to reconstruct the radial and bulk flow as a function of the distance. The method was applied to the Cosmicflows-4 (CF4) dataset after being tested on a simulated mock data. From the analysis of the radial flow in a sphere of $160\Mpch$, augmented by a simple assumption on the shape of the velocity field on large scales, we provided an estimate of the Hubble parameter $H_0$. 

The application of our method to a series of 64 realistic mock observations of the CF4 catalog yielded two results. First, we quantified the selection effects in space and magnitude on the radial and bulk flow. Although no systematic shift appears, the radial flow and the Cartesian components of the bulk flow are amplified with respect to the underlying signal, as previously discussed in \citet{Andersen2016a}. This results in an overestimated amplitude of the bulk flow, with a qualitative degradation of the reconstructed amplitude beyond $140\Mpch$. Second, a simple simulation-based correction was extracted from the application of our method to the 64 mock catalogs. This approach enabled a corrected estimation of the radial and bulk flow of the CF4 data at all distances up to the edge of the data. 

The uncorrected bulk flow differs significantly from \LCDM{} beyond $140\Mpch$ in the $-$SGX direction and on the magnitude of the bulk flow. The tension is extreme: the bulk flow measured naively rules out \LCDM{}. The increase in the bulk flow around $100-200\Mpch$ has been consistently described in the literature, with various amplitudes. Beyond this distance, the amplitude predicted by the uncorrected approach stagnates, while the one predicted by the corrected estimator decreases slowly. We note that our uncorrected estimation of the bulk flow of CF4 is consistent with that of \citet{Whitford2023} and \citet{Watkins2023}, which might indicate that their methods are also sensitive to spurious amplification issues. However, our studies of the effect of the CF4 survey footprint (more specifically, the angular and radial selection function) have demonstrated that the method overestimates the true velocity field. When this is taken into account, the results are more consistent with \LCDM. This does not need to be the case: our \LCDM{} based correction does not force agreement between the method and the model. It simply indicates that if we lived in a \LCDM{} model, the bulk flow estimated by our forward model would overestimate the true one by a quantifiable amount.

The radial flow in spheres is degenerate with the value of $H_0$ chosen when running the fitting procedure. Instead of exploring the tension with \LCDM{} using the $H_0$ provided by \citet{Tully2023}, we chose to estimate $H_0$ in the light of the CF4 data in a probabilistic way, assigning each value of $H_0$ a probability based on the likelihood of that corresponding mean inflow in \LCDM{}. This analysis resulted in a value of $H_0=75.9\pm1\Hdim$, which is at the higher end of the range of local values of $H_0$ found by other authors. We note that systematic errors on the value should be treated more carefully to obtain a more realistic uncertainty range. We also highlight that this estimate differs from a crude regression of the $cz = H_0 d$ on the data, as in \citet{Tully2023}. In the presence of an anisotropic dataset, the radial and bulk flow are correlated, and the bulk flow thus biases the estimation of $H_0$. Our procedure extracts the radial component from the signal, resulting in a value of $H_0$ that is not polluted by the directional bulk flow. 

Throughout this work, we tried to control the effect of the assumed cosmology on the reported results. 
\begin{itemize}
    \item To reconstruct the radial and bulk flows, an isotropic $H_0$ was assumed, as well as the existence of a velocity field. No prior on the power spectrum of this field was assumed, however, as opposed to previous works \citep[\eg{}][]{Courtois2023, Valade2024, Hoffman2024}.
    \item To derive the distance-redshift relation in \cref{eq:dcosint}, a flat \LCDM{} Universe was considered. Any cosmology assuming an FLRW metric \citep[\eg{}][]{Robertson1933} yields a similar relation, however. 
    \item Our series of synthetic observations of the CF4 catalog and the resulting correction were based on \LCDM{} simulations. The departure of the reconstructed velocity profile with respect to the ground truth is essentially due to the selection function of the data and is thus a priori unrelated to the simulation cosmology. A series of tests that we do not present here demonstrated, however, that the correction scheme depends on cosmological parameters such as $f\sigma_8$. We would like to highlight that assuming \LCDM{} in the correction model is consistent with our aim to test \LCDM{} as a cosmological model. Furthermore, the correction works in favor of \LCDM{} and might only dampen the tension in the data. 
    \item In order to give an estimate of $H_0$, we assumed the cosmological principle, \ie{} that the radial velocities of spheres of increasing radii asymptotically converge to zero. Furthermore, a single-parameter scatter on the radial flow of a sphere of $160\Mpch$ was introduced to model the local cosmic variance. The numerical value of this scatter was extracted from our \LCDM{} simulations. We highlight, however, that (1) \LCDM{} agrees with linear theory on these scales, making our chosen value more general than \LCDM{}; (2) that this scatter only affects the uncertainty on $H_0$ and not its value, and (3) that this result can easily be extended to any other cosmology, given that the cosmic variance on the radial velocity of sphere of $160\Mpch$ can be derived.
    \item It appears that the bulk flow reconstructed by our algorithm is mildly sensitive to the value of $H_0$. Considering, for instance, Planck's $H_0 = 67.77 \Hdim$ value slightly decreases the magnitude of the bulk flow, but simultaneously induces extreme outflows (on the order of thousands of km/s) at the edge of the data. This only marginally mitigates the tension on the amplitude of the bulk flow while creating a major (unphysical) radial outflow of matter. 
\end{itemize} 

This method differs from traditional bulk flow estimators, such as those presented in \citet{Whitford2023}, from several perspectives. First, the forward-modeling approach enables the writing of complex, nonlinear, but more realistic observational models \citep[\eg{}][]{Boubel2024}, thus bypassing external Gaussianization methods \citep[\eg{}][]{Watkins2015, Qin2021, Hoffman2021}. The fitting of the forward model in shells that are combined in the postprocessing into spheres constitutes a new weighting technique. While a simple cut in the data at the high-redshift end causes all the profiles to plateau to their values at small distances because the information is concentrated near the observer, this approach enables a better estimation of the evolution of the fitted values with distance. Compared to field-level inference approaches \citep[\eg{}][]{Courtois2023, Hoffman2024}, our method does not rely on the linear theory of fields, nor does it assume a correlation function of the velocity field, which is known to dampen the tension between the data and the underlying model. Last but not least, our approach is similar to that of \citet{Boubel2024, Boubel2025}. However, we treated the entire CF4 catalog with a more general, but less realistic, model than did \citet{Boubel2024}, who focused on the Tully-Fisher relation and were thus limited to about $10\,000$ constraints out of the $38\,000$ of the entire CF4 grouped catalog. While interpreting the dipolar anisotropic expansion as bulk flow, \citet{Kalbouneh2025b} reported similar bulk flow results using the ungrouped CF4 data, complementing the findings of \citet{Stiskalek2025} using the redshift-limited $(z\lesssim0.05)$ CF4 subsamples, where no local \Hnod anisotropy was detected, and the results remain consistent with \LCDM{} at these shallower scales.

In future work, we suggest improving the model to include possible systematics in the observational data, notably issues of selection function or intercalibration, which have been proven to lead to significant spurious flows \citep[Fig. 9]{Whitford2023}. The correction framework could be improved by considering more elaborated methods. We also look forward to more peculiar velocity data, specifically, to data covering the heavens more spherically.

\begin{acknowledgements}

CD appreciates the DPST scholarship. AV thanks A. Whitford for sharing the literature data used in \cref{fig:summary}. AV has been supported by the Agence Nationale de la Recherche of the French government through the program ANR-21-CE31-0016-03. NIL acknowledges funding from the European Union Horizon Europe research and innovation program (EXCOSM, grant No.101159513)/ The MultiDark Database used in this paper and the web application providing online access to it were constructed as part of the activities of the German Astrophysical Virtual Observatory as a result of a collaboration between the Leibniz-Institute for Astrophysics Potsdam (AIP) and the Spanish MultiDark Consolider Project CSD2009-00064. The Bolshoi and MultiDark simulations were run on the NASA’s Pleiades supercomputer at the NASA Ames Research Center. The MultiDark-Planck (MDPL) and the BigMD simulation suite have been performed in the Supermuc supercomputer at LRZ using time granted by PRACE. Plots are made with Python libraries: \textit{Healpy} and \textit{Matplotlib}. HMC module is made available by \textit{TensorFlow Probability}, coupled with \textit{NumPy}, \textit{Scipy}, and \textit{Astropy} for handy mathematical constants and expressions. 
\end{acknowledgements}

\bibliographystyle{bibtex/aa}
\bibliography{main}

\begin{appendix}
\section{Modeling the magnitude-limited mocks}\label{app:magcut}

The mocks presented in this work are based on halos issued from a dark matter-only simulation. In order to properly model the magnitude cut, the halos have to be affected a luminosity. This is done by introducing a simplistic mass-luminosity relationship
\begin{equation}
    \label{eq:lm_rel}
    \frac{L}{L_0} = \alpha \left(\frac{M}{M_0}\right)^\beta
\end{equation}
where $L$ is the constructed luminosity, $M$ is the tabulated mass of the halos (using the $200c$ convention) and $L_0=10^{10}$ and $M_0=10^{12}$ are normalizing constants. Given the halo mass function $P\big(\log(M)\big)$  predicted by \citet{Tinker2008}, the expected number of halos as function of distance in the presence of a magnitude cut at $m_{\rm cut}$ is given by
\begin{equation}
    \begin{aligned}
        P^{\rm mod}(r) = &\,\, r^2 \times \delta^{K}(r < r_{\rm cut}) \\ 
                         &\times \int_{- \infty}^{m_{\rm cut}} P\left(\log(M) = \beta^{-1} (m - 5 \log(r) - 25) \right) {\rm d} m
    \end{aligned}
\end{equation}
where $\delta^K$ is the Kronecker delta function and $\delta^{K}(r < r_{\rm cut})$ is a crude modeling of the cut in redshift space present in the 6dF and SDSS catalogs. Note that we have set $\alpha = 1$ as this parameter is perfectly degenerated with the parameter $m_{\rm cut}$. 

The parameters $\beta$ and $m_{\rm cut}$ are fitted to each one of the major subcomponents of the CF4 catalog. The optimized quantity is 
\begin{equation}
    \chi^2 = \sum_{i} \left( P^{\rm obs}(r_i) - P^{\rm mod}(r_i) \right)^2 
\end{equation}
where $r_i = i \times \Delta r$ with $\Delta r = 5\Mpch$ and $P^{\rm obs}(r)$ is the histogram of the CF4 redshift-derived distances. 

The results of the fitting procedure are presented in \cref{tab:mocks} and in \cref{fig:mock_spat_dist_type_1} (right panel). The mass-distance relationship, displaying the simulated magnitude cut, is presented in \cref{fig:apd_magcut_distr} in the appendix. A visual evaluation of the distance histogram shows a good convergence of the approach. The variations in the CF4 observations appear in agreement with the cosmic variance over our 64 mocks, represented by the shaded corridor around the mean histogram. Even though the simplicity of the model prevents a physical interpretation for the fitted values of $m_{\rm cut}$ and $\beta$, it is noted that all the subcatalogs first display a volume-limited regime before being limited in magnitude, at $60\Mpch$ for the \other{} component, $90\Mpch$ for 6dF and $230\Mpch$ for SDSS.

\begin{figure}[!h]
    \centering
    \includegraphics[width=\linewidth]{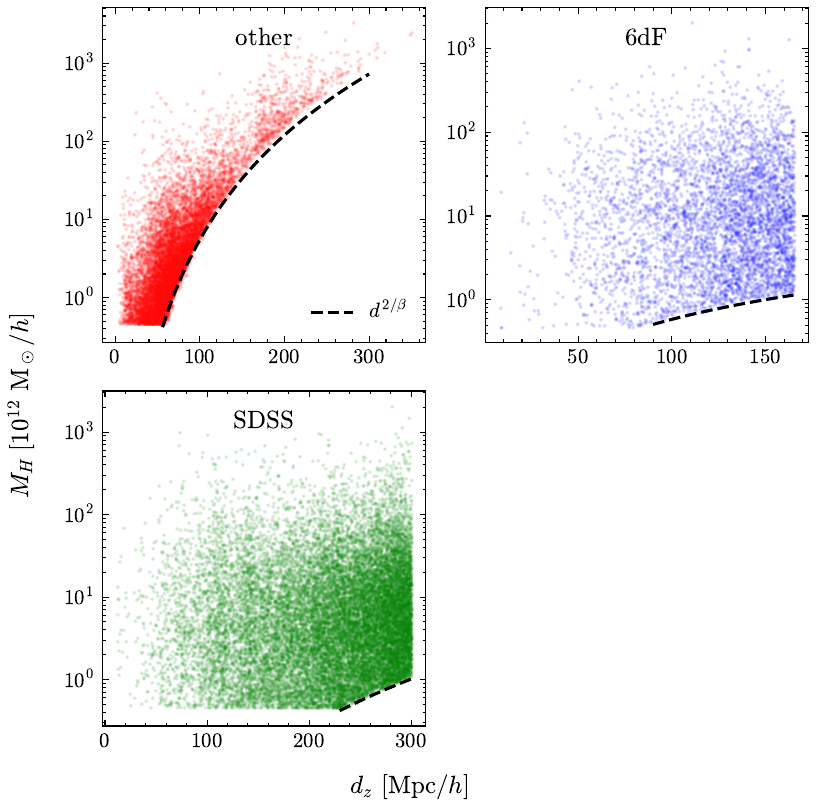}
    \caption{An example of the halo mass as a function of redshift distance of the selected halo in the mock, given the criteria in \cref{tab:mocks}. The minimum halo mass cut (equivalent to the minimum magnitude cut) is in a black dashed line.}
    \label{fig:apd_magcut_distr}
\end{figure}

\section{Effect of $\Omega_M$} \label{app:omega_m}

\begin{figure}[!h]
    \centering
    \includegraphics[width=\hsize]{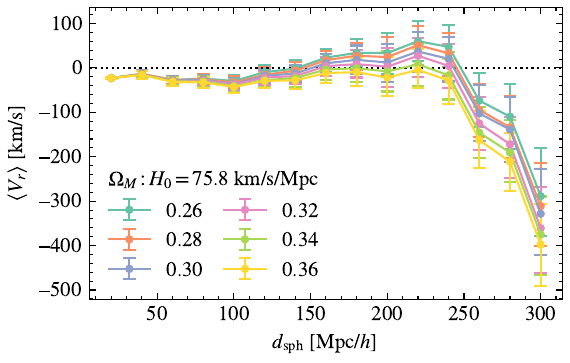}
    \caption{Radial flow fit of CF4 data with different matter density parameter, given the same $H_0$.}
    \label{fig:omega_m}
\end{figure}

\Cref{eq:dcosint} introduces a dependence of our model on $H_0$ and $\Omega_M$. As demonstrated in \cref{fig:omega_m}, the effect of $\Omega_M$ is of second order compared to that of $H_0$. A strong variation of $\Omega_M$ ($\pm3\sigma$ the uncertainty of \citet{Brout2022}) results in a variation of less than $100\kms$ in the radial flow at $240\Mpch$, comparable to a variation of $H_0$ of about $\approx 0.4 \Hdim$. In future works, this should be added in the systematic errors budget on $H_0$. 

\end{appendix}

\end{document}